\documentclass[prb,twocolumn,floatfix,superscriptaddress]{revtex4}

\usepackage{graphicx}
\usepackage{amsmath}
\usepackage{amssymb}
\usepackage[colorlinks=true,citecolor=blue,linkcolor=blue]{hyperref}
\usepackage{amsfonts}
\usepackage{color,xcolor}
\usepackage{epstopdf}
\usepackage{braket}
\usepackage{bm}

\graphicspath{{figures/}}
\renewcommand{\vec}[1]{\ensuremath{\boldsymbol{#1}}}

\begin{document}

\title{Excitons, trions, and biexcitons in transition metal dichalcogenides: magnetic field dependence}
\date{\today}
\author{M. Van der Donck}
\email{matthias.vanderdonck@uantwerpen.be}
\affiliation{Department of Physics, University of Antwerp, Groenenborgerlaan 171, B-2020 Antwerp, Belgium}
\author{M. Zarenia}
\email{mohammad.zarenia@uantwerpen.be}
\email{zareniam@missouri.edu}
\affiliation{Department of Physics, University of Antwerp, Groenenborgerlaan 171, B-2020 Antwerp, Belgium}
\affiliation{Department of Physics and Astronomy, University of Missouri, Columbia, Missouri 65211, USA}
\author{F. M. Peeters}
\email{francois.peeters@uantwerpen.be}
\affiliation{Department of Physics, University of Antwerp, Groenenborgerlaan 171, B-2020 Antwerp, Belgium}

\begin{abstract}
The influence of a perpendicular magnetic field on the binding energy and structural properties of excitons, trions, and biexcitons in monolayers of semiconducting transition metal dichalcogenides (TMDs) is investigated. The stochastic variational method (SVM) with a correlated Gaussian basis is used to calculate the different properties of these few-particle systems. In addition, we present a simplified variational approach which supports the SVM results for excitons as a function of magnetic field. The exciton diamagnetic shift is compared with recent experimental results and we extend this concept to trions and biexcitons. The effect of a local potential fluctuation, which we model by a circular potential well, on the binding energy of trions and biexcitons is investigated and found to significantly increase the binding of those excitonic complexes.

\end{abstract}

\maketitle

\section{Introduction}
Two dimensional (2D) transition-metal dichalcogenide (TMD) monolayers, such as MoS$_2$, MoSe$_2$, WS$_2$, WSe$_2$, WTe$_2$, etc. \cite{mak1,mak2,zeng,splendiani,cao}, are currently the subject of numerous theoretical and experimental studies. This is due to their remarkable electronic properties such as an intrinsic spin-orbit coupling (SOC) resulting in a splitting of the energy bands with opposite spins\cite{theory1} and, most notably, the fact that inversion symmetry breaking leads to the formation of a direct band gap, as opposed to the gapless spectrum of graphene\cite{novo1,review}, which is located at the two inequivalent $K$ and $K'$ valleys at the corners of the first Brillouin zone. These properties make these materials promising for future electronic and optic applications, as well as for novel valleytronic applications \cite{mak2,zeng,cao}.

The 2D nature of TMD monolayers leads to strongly enhanced Coulomb interactions, which are also influenced by the dielectric environment \cite{mismatch}. This leads to the formation of tightly bound \emph{excitons}, a bound system consisting of an electron and a hole. The binding energy of excitons in these materials can be of the order of 0.5 eV, which is one to two orders of magnitude larger as compared to excitons in conventional semiconductors, which have been investigated for more than half a century \cite{elliot,kulak, francoisE,francoisT,expT0}. Excitonic states were indeed found in the band gap of monolayer TMDs in photoluminescence experiments, both in the absence \cite{mak3,he,sallen,korn} and presence \cite{perpfield,perpabs1,perpabs2,perpabs3,perpabs4,ws2dia,wse2dia} of a perpendicular magnetic field. There have also been a few theoretical studies on the excitonic absorption spectrum of these materials which were limited to zero magnetic field \cite{exctheory1,exctheory2,exctheory3}.

A bound state of an exciton (X) with an additional electron (e) or hole (h) can be formed. Such three-particle states are known as \emph{trions} and can be either positive (X$^+$) or negative (X$^-$) depending on whether an additional hole or electron is bound, respectively. Since the first prediction of trions in bulk semiconductors \cite{lampert}, there have been many theoretical \cite{francoisE,francoisT,munschy} and experimental \cite{expT0,expT1,expT2,mak2,expT4} studies on these excitonic structures in different systems such as e.g. semiconductor quantum wells (for example see Refs. [\onlinecite{francoisE,francoisT,expT0}]). They have also been observed in TMDs in recent photoluminescence experiments on monolayer MoS$_2$ and WSe$_2$ \cite{mak3,lui,perpabs4}. In these experiments, trion binding energies of 20-30 meV were found, which is one to two orders of magnitude larger than the binding energy of trions in GaAs quantum wells, which is typically of the order of 0.5-3 meV depending on the width of the quantum well \cite{qwell}. The trion binding energy for different TMDs was recently calculated by Berkelbach {\it et al.} in the absence of a magnetic field \cite{berkelbach} using a variational solution of the single-band low-energy model.  

In addition to excitons and trions, one might also expect higher-order few-body quasiparticles, such as \emph{biexcitons}. A biexciton is a system consisting of two excitons which are bound together. There exist several theoretical and experimental studies on biexcitons in bulk and 2D semiconductors (see for example Ref. [\onlinecite{francoisB}] and references therein). Stable biexcitons were recently observed in monolayer TMDs \cite{biexciton}. Due to the strong long-range Coulomb interaction in TMD monolayers, a large biexciton binding energy of about 35-60 meV was measured.

Previous theoretical studies on excitonic complexes in TMD monolayers are, to our knowledge, limited to the case of zero magnetic field \cite{analytic,analytic2}, which is in contrast to semiconductors where the magnetic field dependence has been thoroughly investigated \cite{qwell,expT0} both theoretically and experimentally. In the present paper we investigate the influence of a perpendicular magnetic field on the exciton, trion, and biexciton binding energy of several monolayer TMD materials. We employ the stochastic variational method (SVM) using a correlated Gaussian basis \cite{svm1,svm2}. This approach was successfully used to describe the binding energy of excitons, trions, and biexcitons in semiconductor quantum wells \cite{francoisE} and their magnetic field dependence \cite{francoisT}. Recently this approach was used to calculate the binding energy of excitons, trions, and biexcitons in TMD monolayers in the absence of a magnetic field \cite{analytic,analytic2} and reasonable agreement with experiments and other theoretical results was found for the exciton and trion binding energies. Here, we will use this approach also to study the effect of a circular potential well on the binding energy of excitonic systems. The motivation of the latter is that such a potential is the simplest model allowing us to study the effect of confinement that may result from local disorder or potential fluctuations\cite{qwell}.

Our paper is organized as follows. In Sec. \ref{sec:Model} we present an outline of the stochastic variational method together with a simplified variational approach for the exciton binding energy as function of an applied magnetic field. The numerical results are discussed in Sec. \ref{sec:Results}. In Sec. \ref{sec:Summary and conclusion} we summarize the main conclusions.

\section{Model}
\label{sec:Model}

\subsection{Stochastic variational method}

\begin{figure}
\centering
\includegraphics[width=8.5cm]{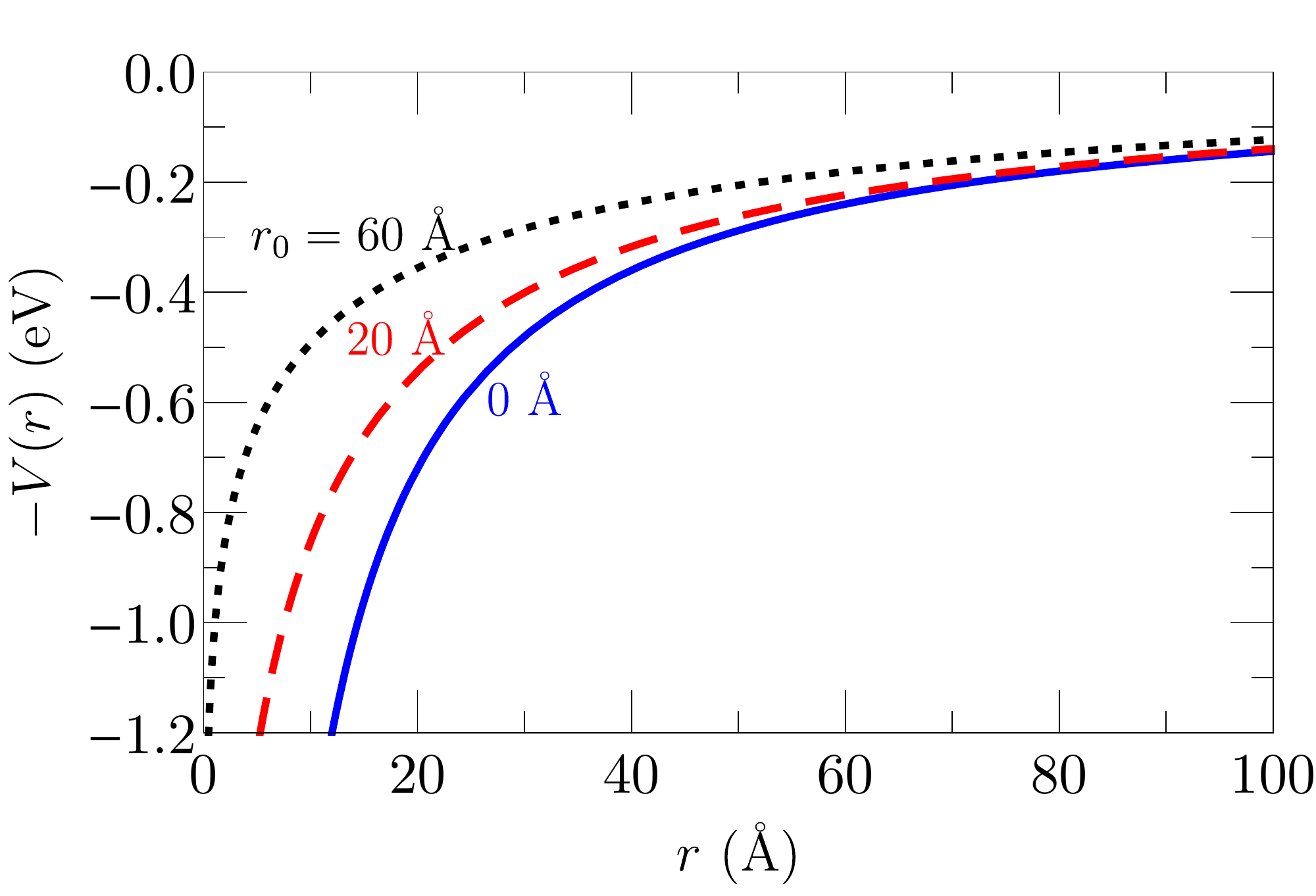}
\caption{(Color online) Interaction potential between a hole and an electron in a TMD suspended in vacuum ($\kappa=1$) with screening length $r_0=0$ \AA\ (solid, blue), $r_0=20$ \AA\ (dashed, red), and $r_0=60$ \AA\ (dotted, black).}
\label{fig:screening}
\end{figure}

The Hamiltonian for an $N$-particle excitonic system in the presence of a magnetic field is given by
\begin{equation}
\label{ham}
H = \sum_{i=1}^N\frac{\hbar^2}{2m_i}\left(\vec{k}_i-\frac{q_i}{\hbar}\vec{A}_i\right)^2+\sum_{i<j}^NV_{ij}(|\vec{r}_i-\vec{r}_j|)+\sum_{i=1}^NV(r_i),
\end{equation}
with $q_i$ and $m_i$ the charge and effective mass of particle $i$. $\displaystyle \vec{A}_i=-\vec{r}_i\times\vec{B}/2$ is the vector potential in the symmetric gauge corresponding to the uniform perpendicular magnetic field $\displaystyle \vec{B}=(0,0,B)$. We do not take into account the different Zeeman terms\cite{perpabs3,perpabs4,valleyzeeman} because they do not influence the binding energy and the structural properties of the excitonic systems. $V(r)=V_0\Theta(r-a_w)$ is the single-particle confinement potential with $V_0$ and $a_w$ the height and radius of the potential well, respectively. For simplicity we assume the well depth to be the same for electrons and holes and that the electron and hole bands are isotropic and parabolic, which is a good approximation for the low energy spectrum of the considered TMDs. The Hamiltonian can be rewritten as
\begin{equation}
\label{ham2}
H = \sum_{i=1}^N\left(V(r_i)-\frac{\hbar^2}{2m_i}\nabla_i^2+\frac{q_i^2B^2}{8m_i}r_i^2-\frac{q_iB}{2m_i}l_{zi}\right)+\sum_{i<j}^NV_{ij},
\end{equation}
with $l_{zi}$ the $z$-component of the angular momentum of particle $i$. The TMD monolayer is surrounded by a dielectric with a dielectric constant different from that of the TMD. Together with the two-dimensional (2D) dielectric screening in the TMD this changes the potential from a $1/r$ Coulomb potential to an interaction potential $V_{ij}$ which is now given by\cite{screening1,screening2,screening3}
\begin{equation}
\label{inter}
V_{ij}= \frac{q_iq_j}{4\pi\kappa\varepsilon_0}\frac{\pi}{2r_0}\left[H_0\left(\frac{|\vec{r}_i-\vec{r}_j|}{r_0}\right)-Y_0\left(\frac{|\vec{r}_i-\vec{r}_j|}{r_0}\right)\right],
\end{equation}
with $Y_0$ and $H_0$ the Bessel function of the second kind and the Struve function, respectively, with $\kappa=(\varepsilon_1+\varepsilon_2)/2$, where $\varepsilon_{1(2)}$ is the dielectric constant of the environment above (below) the TMD monolayer, and with $r_0=2\pi\chi_{2\text{D}}/\kappa$ the screening length with $\chi_{2\text{D}}$ the 2D polarizability of the TMD. In general TMDs are placed on a substrate with a dielectric constant $\varepsilon_2=\varepsilon_r$ and with vacuum on top, i.e. $\varepsilon_1=1$. The interaction potential is shown in Fig. \ref{fig:screening} for different screening lengths. For $r_0=0$ this potential reduces to the bare Coulomb potential $V_{ij}=q_iq_j/(4\pi\kappa\varepsilon_0r_{ij})$ with $r_{ij}=|\vec{r}_i-\vec{r}_j|$. Increasing the screening length leads to a decrease in the short-range interaction strength while the long-range interaction strength is unaffected. For very large screening lengths $r_0\rightarrow\infty$ the divergence in $r=0$ becomes logarithmic, i.e. $V_{ij}=q_iq_j/(4\pi\kappa\varepsilon_0r_0)\text{ln}(r_0/r_{ij})$.

The Schr\"odinger equation for the few-particle system can not be solved exactly. Therefore, in order to calculate the energies of the different excitonic systems described by the above Hamiltonian, we employ the SVM in which the many-particle wave function $\Psi(\vec{r}_1,\ldots,\vec{r}_N)$ is expanded in a basis of size $K$\cite{svm1,svm2}:
\begin{equation}
\label{basisexp}
\Psi_{M_L,S,M_S}(\vec{r}_1,\ldots,\vec{r}_N) = \sum_{n=1}^Kc_n\varphi_{M_L,S,M_S}^{n}(\vec{r}_1,\ldots,\vec{r}_N),
\end{equation}
where the basis functions are taken as correlated Gaussians:
\begin{equation}
\label{corrgaus}
\begin{split}
&\varphi_{M_L,S,M_S}^{n}(\vec{r}_1,\ldots,\vec{r}_N) = \\
&\mathcal{A}\left(\prod_{j=1}^N[x_j+iy_j\text{sgn}(m_j^n)]^{|m_j^n|}e^{-(\vec{x}^TA_n\vec{x}+\vec{y}^TA_n\vec{y})/2}\chi^n_{S,M_S}\right),
\end{split}
\end{equation}
where $\vec{x}$ and $\vec{y}$ are vectors containing the $x$-components $x_j$ and $y$-components $y_j$, respectively, of the different particles. The matrices $A_n$ are symmetric and positive definite and contain variational parameters. $\chi^n_{S,M_S}$ is the total spin state of the excitonic system with total spin $S$ and $z$-component $M_S$. Multiple total spin states belonging to the same $S$ and $M_S$ value are possible since these are obtained by adding step by step single-particle spin states and therefore different intermediate spin states may result in the same total spin state. The integers $m_j^n$ represent possible values of the angular momentum of the different particles and satisfy the relation $\sum_{j=1}^Nm_j^n=M_L$ with $M_L$ the $z$-component of the total angular momentum. Finally, $\mathcal{A}$ is the antisymmetrization operator for the indistinguishable particles. The calculation of the matrix elements of the different terms of the Hamiltonian between these basis functions can be done analytically\cite{analytic}.

The procedure for finding the best energy value is as follows. First, a matrix $A_n$, integers $m_j^n$, and a spin function $\chi^n_{S,M_S}$ are randomly generated multiple times. The set of parameters that gives the wave function which has the lowest variational energy is then retained and defines the first basis function. At this point we have a basis of dimension $K=1$. Next, a set of parameters is again generated randomly multiple times and the variational energy is calculated in the $K=2$ basis consisting of the previously determined basis function and the new trial basis function defined by the new set of parameters. The set of parameters which gives the trial function which has the lowest variational energy is then retained and defines the second basis function. Following this procedure, each addition of a new basis function will lead to a lower variational energy and the basis size is increased until sufficient convergence of the variational energy is reached. Here, we found that, when 150 parameter sets are generated to determine a new basis function, a basis size of $K=50$ for excitons and $K=250$ for trions and biexcitons results in an energy convergence of the order of 0.001 $\mu$eV, 0.1 $\mu$eV, and 1 $\mu$eV for excitons, trions, and biexcitons, respectively. This procedure is explained in more detail in Ref. [\onlinecite{svm1}].

\subsection{Simplified variational method for excitons in a magnetic field}

In addition to the SVM approach and in order to get a better physical insight we present here also a simplified variational method for excitons in a magnetic field. Writing out the Hamiltonian \eqref{ham2} for an exciton with zero angular momentum gives
\begin{equation}
\label{exham}
H_{ex} = \frac{p_h^2}{2m_h}+\frac{p_e^2}{2m_e}+\frac{e^2B^2}{8m_h}r_h^2+\frac{e^2B^2}{8m_e}r_e^2+V_{he}(|\vec{r}_h-\vec{r}_e|).
\end{equation}
Introducing center of mass and relative coordinates we use the substitutions
\begin{equation}
\label{transform}
\begin{split}
&\vec{R} = \frac{m_h\vec{r}_h+m_e\vec{r}_e}{m_h+m_e}, \quad \vec{r} = \vec{r}_h-\vec{r}_e, \\
&\vec{P} = \vec{p}_1+\vec{p}_2, \quad \vec{p} = \frac{m_e\vec{p}_h-m_h\vec{p}_e}{m_h+m_e},
\end{split}
\end{equation}
which leads to the Hamiltonian
\begin{equation}
\label{exham2}
\begin{split}
H_{ex} = &\frac{P^2}{2M}+\frac{e^2B^2}{8\mu}R^2+\frac{p^2}{2\mu}+\frac{e^2B^2}{8M^2}\left(\frac{m_h^3+m_e^3}{m_hm_e}\right)r^2 \\
&+\frac{e^2B^2}{4}\frac{(m_e-m_h)}{M\mu}\vec{R}.\vec{r}+V_{he}(r),
\end{split}
\end{equation}
with $M=m_h+m_e$ and $1/\mu=1/m_h+1/m_e$. For equal electron and hole masses $m_e=m_h=m$, this decouples into a center of mass part and a relative part, i.e. $H_{ex}=H_{CM}+H_{rel}$, $\Psi_{ex}(\vec{R},\vec{r})=\psi_{CM}(\vec{R})\psi_{rel}(\vec{r})$, and $E_{ex}=E_{CM}+E_{rel}$. The center of mass part can be solved exactly. We can rewrite the Hamiltonian as
\begin{equation}
\label{mcham}
H_{CM} = \frac{P^2}{2M}+\frac{1}{2}M\omega^2R^2,
\end{equation}
with $\omega=eB/(2\sqrt{M\mu})=eB/M$. This is the Hamiltonian of the 2D harmonic oscillator, which has an energy spectrum given by $E_{CM}=\hbar\omega(n_x+n_y+1)$ with $n_x$ and $n_y$ quantum numbers, yielding a ground state energy of $E_{CM}^0=\hbar eB/M = \hbar^2/(2ml_B^2)$ with $l_B=\sqrt{\hbar/eB}$ the magnetic length. The corresponding ground state wave function is given by
\begin{equation}
\label{mcgolf}
\psi_{CM}^0(\vec{R}) = \frac{1}{\sqrt{\pi}l_B}e^{-R^2/(2l_B^2)}.
\end{equation}
Note that this implies that a difference in electron and hole mass would only lead to corrections of the order of $(m_e-m_h)^2$ since first order perturbation theory implies that the lowest order correction of the corresponding term in the Hamiltonian is proportional to $\braket{\psi_{CM}^0|R|\psi_{CM}^0}=0$.

The relative part of the Hamiltonian can be written as
\begin{equation}
\label{relham}
H_{rel} = -\frac{\hbar^2}{2\mu}\nabla^2+\frac{e^2B^2}{32\mu}r^2+V_{he}(r).
\end{equation}
In the case of zero magnetic field the Hamiltonian reduces to that of a hydrogen-like problem which, in the absence of screening, has an exponential ground state wave function. Without the Coulomb-like interaction term, on the other hand, the Hamiltonian reduces to that of an harmonic oscillator which has a Gaussian ground state wave function. Therefore, to interpolate between these limiting cases, we consider the following variational wave function
\begin{equation}
\label{relgolf}
\psi_{rel}^0(\vec{r}) = Ne^{-a^2r^2-br},
\end{equation}
with $a$ and $b$ variational parameters and $N$ a normalization constant. The variational ground state energy of the relative part of the Hamiltonian is
\begin{equation}
\label{relener}
E_{rel}^0(a,b) = \frac{\braket{\psi_{rel}^0|H_{rel}|\psi_{rel}^0}}{\braket{\psi_{rel}^0|\psi_{rel}^0}},
\end{equation}
and the best approximation for the total exciton energy is therefore given by
\begin{equation}
\label{Etot}
E_{ex} = \frac{\hbar^2}{2ml_B^2}+E_{rel}^0(a_{min},b_{min}),
\end{equation}
with $a_{min}$ and $b_{min}$ the variational parameters which minimize the variational energy.

\subsection{Relevant quantities}

We calculate the binding energies for excitons, negative trions, and biexcitons, which are, respectively, given by
\begin{align}
\label{binding}
&E_b^{ex}(B,r_0) = E_0^e(B)+E_0^h(B)-E_{ex}(B,r_0), \\
&E_b^{tr}(B,r_0) = E_0^e(B)+E_{ex}(B,r_0)-E_{tr}(B,r_0), \\
&E_b^{bi}(B,r_0) = 2E_{ex}(B,r_0)-E_{bi}(B,r_0),
\end{align}
where $E_0^{e(h)}$, $E_{ex}$, $E_{tr}$ and $E_{bi}$ are the free electron (hole), exciton, trion, and biexciton energy, respectively.

Furthermore, the correlation function between two particles $i$ and $j$, is defined as
\begin{equation}
\label{corr}
C_{ij}(\vec{r}) = \braket{\Psi|\delta(\vec{r}_i-\vec{r}_j-\vec{r})|\Psi},
\end{equation}
from which we can calculate the probability of finding particles $i$ and $j$ at a distance $r$, which for an axial symmetric system reduces to
\begin{equation}
\label{prob}
P_{ij}(r) = 2\pi rC_{ij}(r),
\end{equation}
which satisfies
\begin{equation}
\label{norm}
\int_0^{\infty}P_{ij}(r)dr = 1.
\end{equation}
The average distance between particles $i$ and $j$ is then obtained by
\begin{equation}
\label{dist}
\braket{r_{ij}} = \int_0^{\infty}rP_{ij}(r)dr = 2\pi\int_0^{\infty}r^2C_{ij}(r)dr.
\end{equation}
In the simplified variational model for the exciton one can show that
\begin{equation}
\label{corrsimple}
C_{eh}(r) = \frac{1}{2\pi}\frac{8a^3}{2a-\gamma b}e^{-2a^2r^2-2br},
\end{equation}
leading to an average electron-hole distance given by
\begin{equation}
\label{distsimple}
\braket{r_{eh}} = \frac{\gamma(a^2+b^2)-2ab}{2a^2\left(2a-\gamma b\right)},
\end{equation}
with $\gamma=\sqrt{2\pi}e^{b^2/(2a^2)}\text{Erfc}(b/(\sqrt{2}a))$ and where the magnetic field and screening length dependence is reflected in the variational parameters $a$ and $b$ which have to be chosen such that they minimize the variational energy.

The center of mass part of the Hamiltonian leads to a linear magnetic field term in the exciton energy spectrum \eqref{Etot}. It can be shown that this term is in general given by $N\hbar^2/(2Ml_B^2)$ for an $N$-particle excitonic system with equal effective electron and hole masses. The quadratic part of the excitonic energy spectrum, the so-called diamagnetic shift\cite{diamagnetic}, is approximately given by $\sigma=e^2\braket{r^2}/32\mu$, where the expectation value is taken with respect to the wave function in the absence of a magnetic field. It is possible to show that, up to first order in the electron-hole mass difference, the center of mass part can be decoupled from the relative part of the Hamiltonian and that the diamagnetic shift of the energy of the $N$-particle excitonic system is in general given by
\begin{equation}
\label{diashift}
\sigma = \frac{e^2}{8M}\sum_{i>j}^N\braket{r_{ij}^2}.
\end{equation}
This value can be experimentally obtained by fitting the results of the transition energy as a function of the magnetic field and can as such give information about the size of the excitonic system. The transition energy is defined as the energy of the photon resulting from the recombination process of an electron and a hole in the excitonic system\cite{francoisT}, which gives
\begin{align}
\label{transition}
&E_t^{ex}(B,r_0) = E_g+E_{ex}(B,r_0), \\
&E_t^{tr}(B,r_0) = E_g+E_{tr}(B,r_0)-E_0^e(B), \\
&E_t^{bi}(B,r_0) = E_g+E_{bi}(B,r_0)-E_{ex}(B,r_0),
\end{align}
with $E_g$ the band gap. Since the diamagnetic shift describes the quadratic dependence on the magnetic field and since $E_g$ and $E_0^e(B)$ are, respectively, constant and linear as a function of the magnetic field it follows that $\sigma_t^{ex}=\sigma^{ex}$, $\sigma_t^{tr}=\sigma^{tr}$, and $\sigma_t^{bi}=\sigma^{bi}-\sigma^{ex}$. Therefore, by measuring and fitting the transition energy of a given excitonic system $\sigma_t$ can be obtained, from which in turn $\sigma$ can be found which then gives an estimate of the size of the excitonic system through Eq. \eqref{diashift}. The different Zeeman terms, which are not taken into account in this paper, are linear as a function of the magnetic field and therefore do not influence the diamagnetic shift.

\begin{figure}
\centering
\includegraphics[width=8.5cm]{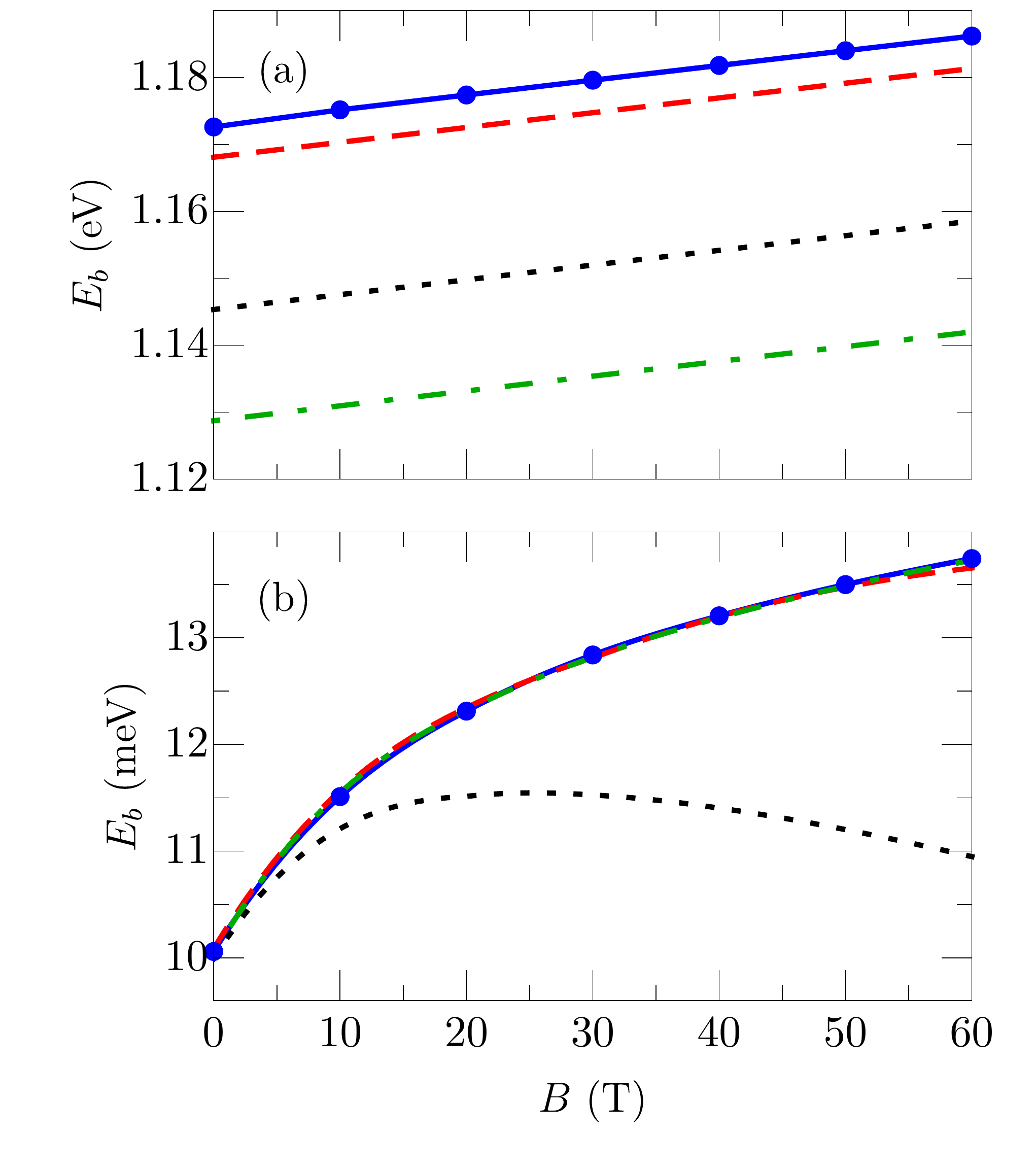}
\caption{(Color online) Exciton binding energy as a function of perpendicular magnetic field for screening length $r_0=1$ nm (a) and $r_0=500$ nm (b). The blue, full curves are obtained with the SVM, the red, dashed curves are the results from the simplified variational model \eqref{relgolf}, the black, dotted curves are obtained with the simplified model in which $a$ is set to 0, and the green, dot-dashed curves are obtained with the simplified model in which $b$ is set to 0.}
\label{fig:excbind}
\end{figure}

\section{Numerical Results}
\label{sec:Results}

We are interested in the ground state and we therefore consider the $(S,M_S)=(0,0)$ singlet state for the exciton and the biexciton and we take the $(S,M_S)=(1/2,1/2)$ doublet state for the trion, while $M_L=0$ is assumed in all three cases.

\subsection{Exciton}

In order to understand the physics we first consider a TMD monolayer suspended in vacuum ($\kappa=1$) and for the electron and hole band masses we assume $m_e=m_h=0.26m_0$ which results in a reduced mass of $\mu=0.13m_0$ with $m_0$ the free electron mass. In Fig. \ref{fig:excbind} we show the exciton binding energy as a function of magnetic field for two different screening lengths. We see that an increased screening length leads to a decreased binding energy, which is a consequence of the decreased short-range interactions, as shown in Fig. \ref{fig:screening}. For small screening lengths, the binding energy increases linearly with the magnetic field strength, whereas for large screening lengths the binding energy initially increases linearly with the magnetic field strength but at higher magnetic field strengths the increase becomes slower than linear. This is because for small screening lengths the linear term in the exciton binding energy dominates over the diamagnetic term in the shown magnetic field range. For large screening lengths, and therefore weak interactions, the diamagnetic term is larger and leads to a deviation from the linear behavior. This deviation starts at lower magnetic field strengths for larger screening lengths. This can also be understood since a perpendicular magnetic field leads to in-plane confinement of the particles, for which the length scale is the magnetic length $l_B=\sqrt{\hbar/eB}$. At low magnetic field strengths this length scale is much larger than the average interparticle distance and this confinement leads to an increase in the binding energy. As the magnetic field strength increases, the magnetic length decreases and eventually becomes of the same order as the interparticle distance. Increasing the magnetic field strength even further will cause the particles to be pushed closer towards each other which increases the kinetic energy. This effect decreases the binding energy and adds up with the increase in binding energy stemming from the magnetic confinement to yield a deviation from the linear magnetic field dependence of the binding energy. At larger screening lengths the interparticle distance is larger due to the decreased Coulomb interactions and therefore the deviation from the linear behavior starts at lower magnetic field strengths.
\begin{figure}
\centering
\includegraphics[width=8.5cm]{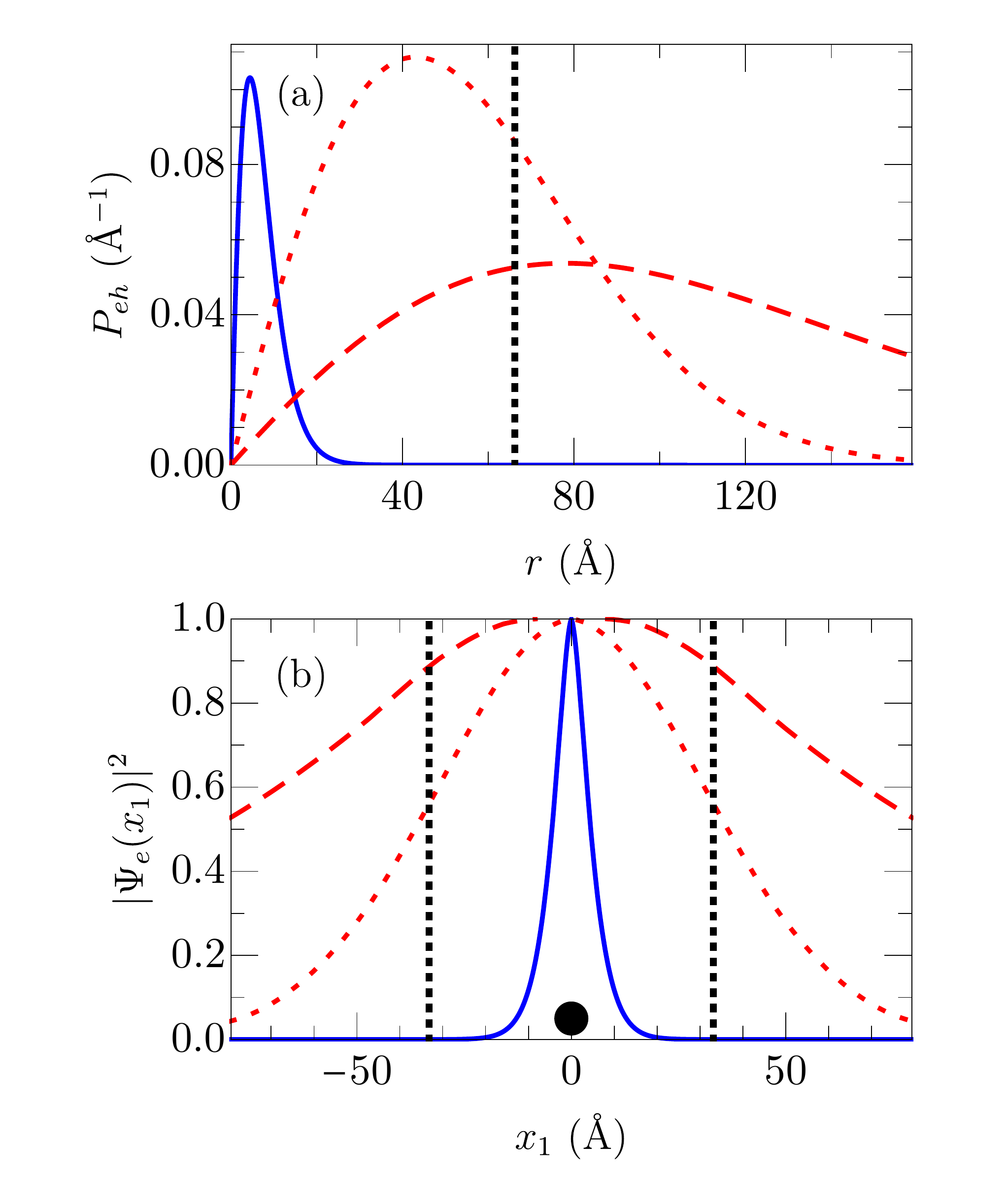}
\caption{(Color online) (a) Interparticle distance probability distribution between the electron and hole of an exciton for screening length $r_0=1$ nm (solid, blue) and $r_0=500$ nm (red, scaled up by a factor of 8), with (dotted) and without (dashed) a perpendicular magnetic field of $B=60$ T. The black dotted vertical line indicates twice the magnetic length. (b) The modulus squared of the wave function $\Psi_e(x_1)\equiv\Psi_{0,0,0}\left((x_1,y_1=0),\vec{0}\right)$ for a fixed hole position indicated by the black dot for the same cases as in (a). The wave functions are rescaled relative to their respective maxima.}
\label{fig:exccorrgolf}
\end{figure}
\begin{figure}
\centering
\includegraphics[width=8.5cm]{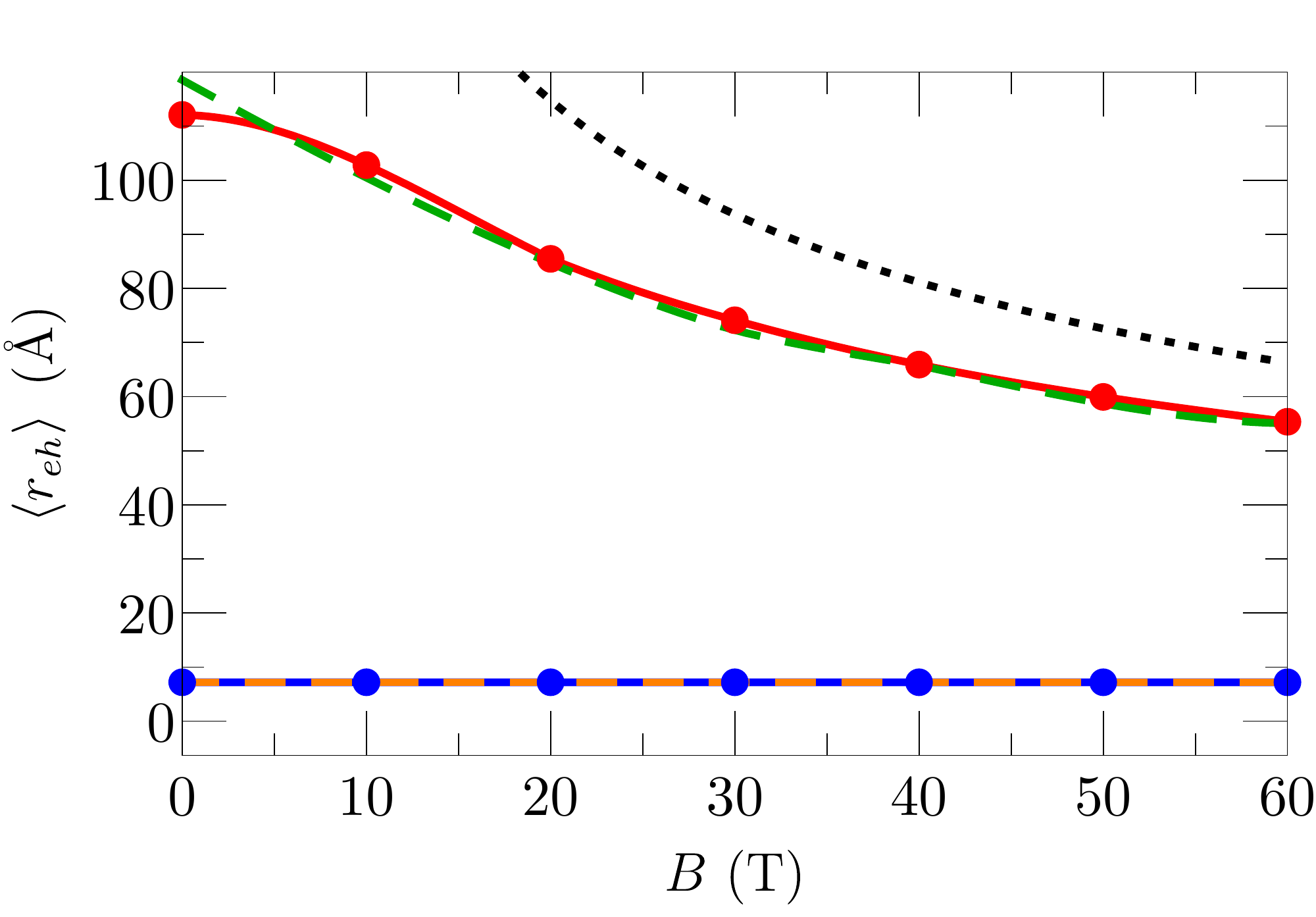}
\caption{(Color online) Average interparticle distance between the electron and hole of an exciton as a function of perpendicular magnetic field for screening length $r_0=1$ nm (SVM: dashed blue and simplified model: dashed orange) and $r_0=500$ nm (SVM: solid red and simplified model: dashed green). The black dotted curve is twice the magnetic length.}
\label{fig:excdist}
\end{figure}

Figure \ref{fig:excbind} also shows that the results obtained with the simplified variational method agree well with those obtained with the SVM. In the absence of screening, which is not shown here, we find that the SVM results can be reproduced with high accuracy in the chosen magnetic field range, i.e. 0 T up to 60 T, by using an exponential variational wave function (i.e. $a=0$). If we use a Gaussian wave function (i.e. $b=0$) we find binding energies that are $21.5\%$ smaller than the SVM values. This implies that in this magnetic field range the Coulomb term dominates over the magnetic field term and therefore the relative part of the exciton is described by an exponential wave function. For small screening lengths Fig. \ref{fig:excbind}(a) shows that the results obtained using the full variational wave function are about 5 meV, or $0.4\%$, smaller than the SVM results, that the results obtained using an exponential wave function ($a=0$) are about 28 meV, or $2.3\%$, smaller than the SVM results, and that the results obtained using a Gaussian wave function ($b=0$) are about 39 meV, or $3.4\%$, smaller than the SVM results. This implies that the interaction term, which is now given by the Keldysh potential of Eq. \eqref{inter} instead of the bare Coulomb potential, still dominates over the magnetic field term. However, the corresponding state can not be described by an exponential or a Gaussian wave function or even a product of the two, although the latter gives the best approximation. In the presence of strong screening Fig. \ref{fig:excbind}(b) shows that the SVM results can be reproduced with high accuracy by using a Gaussian variational wave function ($b=0$). Using an exponential wave function ($a=0$), however, the results agree at low magnetic field strengths but deviate from the SVM results for higher magnetic field strengths, even resulting in a decrease in binding energy. This implies that, due to the strong screening and therefore weak interactions, the magnetic field term now dominates over the interaction term and therefore the relative part of the exciton is described by a Gaussian wave function.

This becomes more clear in Fig. \ref{fig:exccorrgolf}, where we show the interparticle distance probability distribution and the modulus squared of the wave function for a hole fixed at $x=y=0$. When the screening is small these quantities are unaffected by the presence of a magnetic field since the exciton is localized to a region smaller than the magnetic confinement region. In the presence of large screening, however, the exciton is larger than this magnetic confinement region and therefore becomes compressed when a magnetic field is applied. In Fig. \ref{fig:excdist} we show the average interparticle distance as a function of magnetic field. For larger screening lengths, the exciton is larger, which is again a consequence of the decreased interaction. As the magnetic field increases, the exciton in the presence of large screening decreases considerably in size. This is because the magnetic length already becomes comparable to the size of the exciton at a relatively small magnetic field strength of 10 T. Moreover, the average interparticle distance converges to twice the magnetic length for high magnetic field strengths. The size of the exciton in the presence of small screening remains constant at 7.2 \AA\ because it is significantly smaller than the magnetic length in the considered magnetic field strength. The figure also shows good agreement between the SVM and the simplified variational model.
\begin{figure}
\centering
\includegraphics[width=8.5cm]{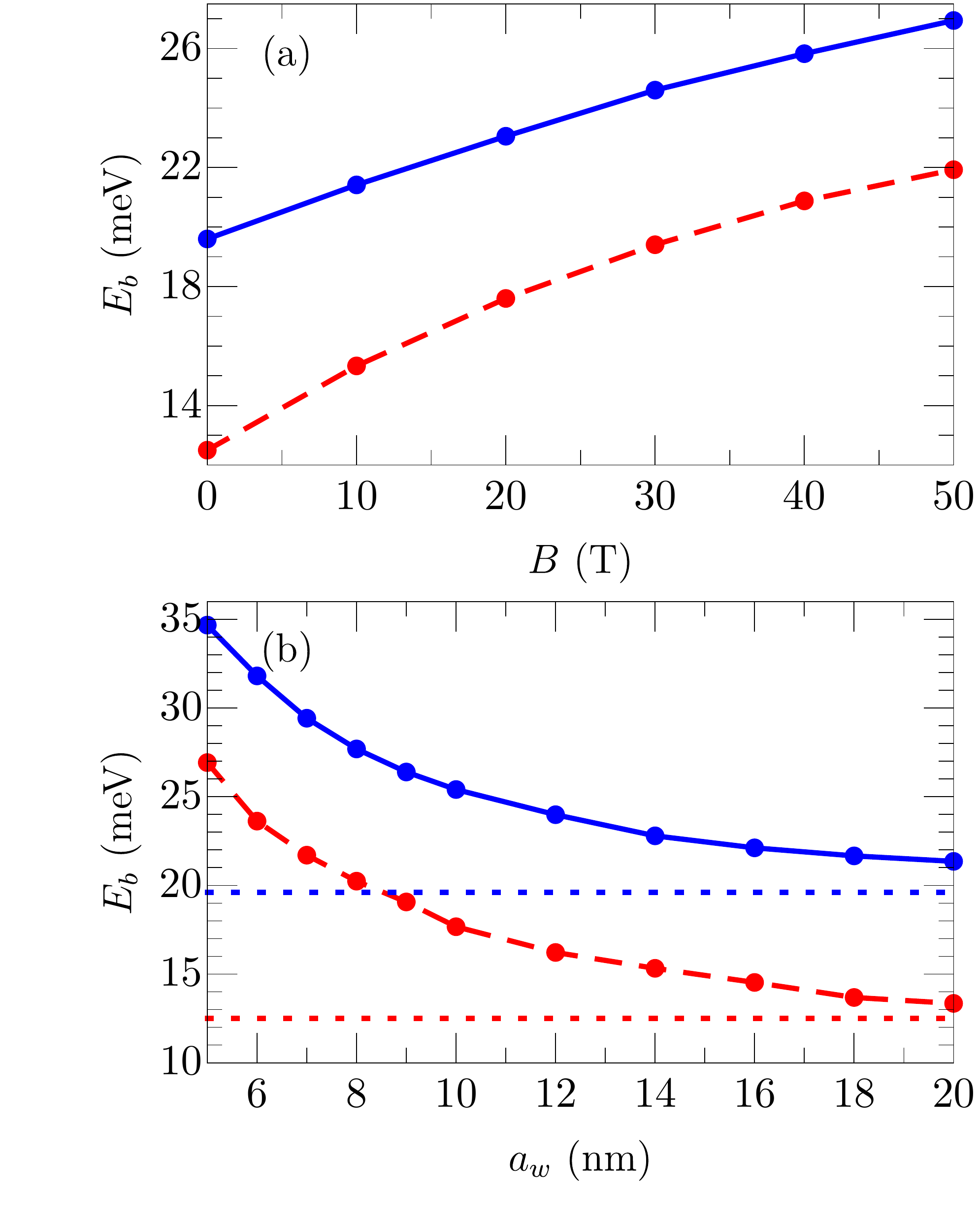}
\caption{(Color online) Negative trion (solid, blue) and biexciton (dashed, red) binding energy as a function of perpendicular magnetic field (a) and potential well radius (b) for WSe$_2$ on a SiO$_2$ substrate. The height of the circular potential well is $V_0=300$ meV. The dotted lines indicate the binding energy in the absence of a potential well.}
\label{fig:plotcomp}
\end{figure}
\begin{figure}
\centering
\includegraphics[width=8.5cm]{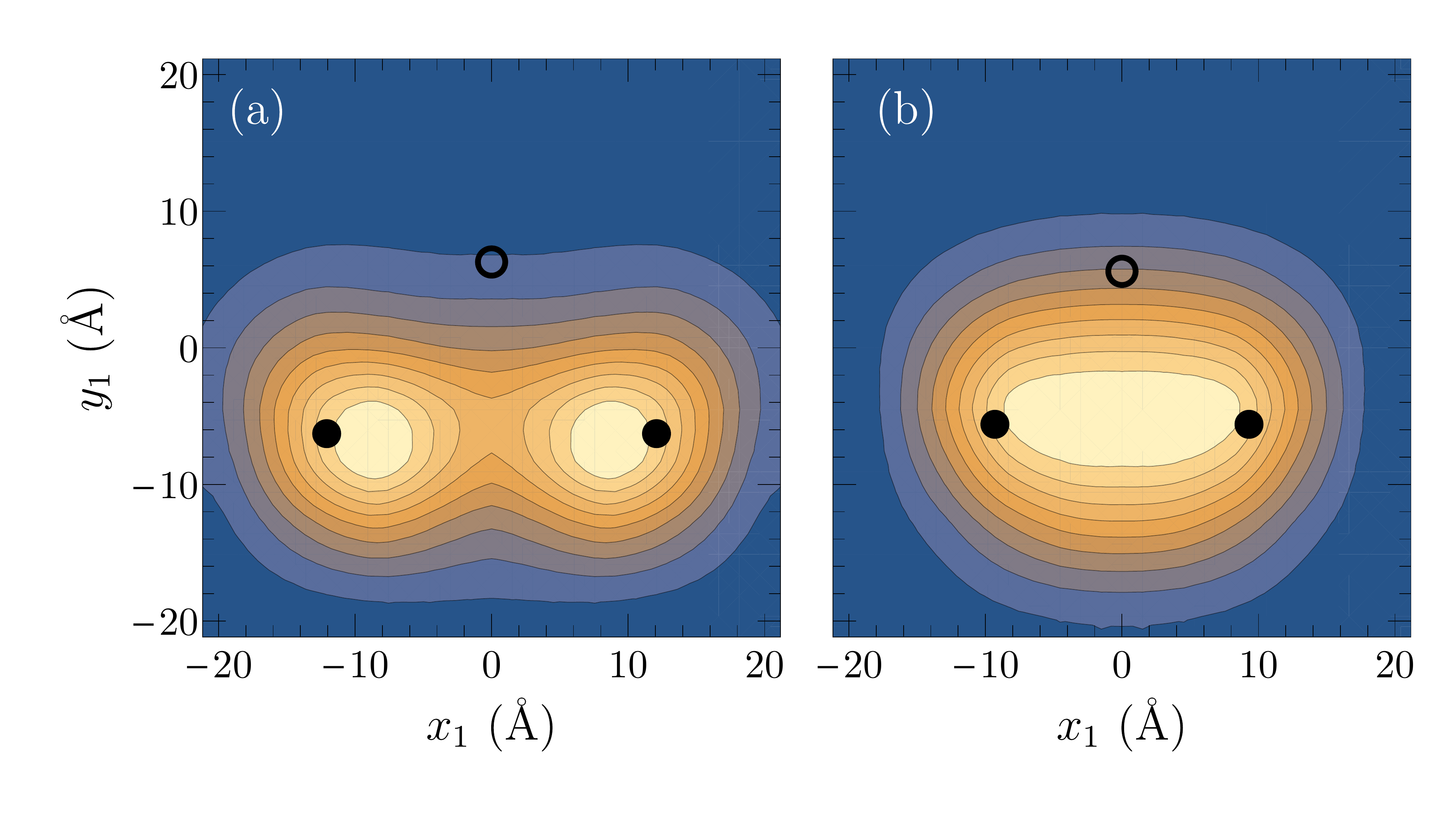}
\caption{(Color online) Modulus squared of the biexciton wave function $\Psi_{0,0,0}\left((x_1,y_1),\vec{r}_2^0,\vec{r}_3^0,\vec{r}_4^0\right)$ for screening length $r_0=26.5$ \AA\ and perpendicular magnetic field $B=0$ T (a) and $B=60$ T (b). The black circles and dots indicate the location of the electrons and holes, respectively.}
\label{fig:bigolfcont}
\end{figure}

\subsection{Trion}

In Fig. \ref{fig:plotcomp}(a) we show the negative trion binding energy as a function of magnetic field for WSe$_2$ on SiO$_2$ substrate, for which we used the parameters given in Table \ref{table:mattable} and a substrate dielectric constant of $\varepsilon_r=3.8$. The behavior is similar to that of the exciton binding energy, however the deviation from the linear behavior starts at higher magnetic field strengths, which is because the corresponding trion interparticle distances are smaller as compared to the magnetic length for the parameters used in this figure. We find $\braket{r_{ee}}=35.6$ \AA\ and $\braket{r_{eh}}=22.5$ \AA, approximately independent of the magnetic field for realistic field strengths, i.e. $B < 100$ T. In Fig. \ref{fig:plotcomp}(b) we show the $B=0$ T negative trion binding energy as a function of the potential well radius for the same material and substrate. The binding energy increases with decreasing radius due to the confinement. For a radius of 5 nm the increase in binding energy is equivalent to what is achievable using a magnetic field of more than 100 T (60 T) for trions (biexcitons).

\subsection{Biexciton}

We show the same results in Figs. \ref{fig:plotcomp}(c)-(d) but now for biexcitons. The biexciton binding energy shows qualitatively the same behavior as the trion binding energy, however the binding energies are smaller than the corresponding trion binding energies. We find $\braket{r_{ee}}=27.8$ \AA\ and $\braket{r_{eh}}=20.5$ \AA, approximately independent of the magnetic field for realistic field strengths, which means that the biexciton is smaller than the trion.
\begin{table}
\centering
\caption{Charge carrier masses, taken from Ref. [\onlinecite{theory2}], and screening lengths, taken from Ref. [\onlinecite{berkelbach}], for different TMD materials suspended in vacuum.}
\begin{tabular}{c c c c}
\hline
\hline
 & $m_e$ ($m_0$) & $m_h$ ($m_0$) & $r_0$ (\AA) \\
\hline
\hline
Mo$\text{S}_2$ & 0.47 & 0.54 & 41.47 \\
\hline
MoS$\text{e}_2$ & 0.55 & 0.59 & 51.71 \\
\hline
W$\text{S}_2$ & 0.32 & 0.35 & 37.89 \\
\hline
WS$\text{e}_2$ & 0.34 & 0.36 & 45.11 \\
\hline
\hline
\end{tabular}
\label{table:mattable}
\end{table}
\begin{table}
\centering
\caption{Exciton binding energies (meV) for different TMD materials, compared with previous $B=0$ T theoretical and experimental studies. The two right columns give our estimates for the magnetic field dependence. We used $\varepsilon_r=4.5$ for hBN (both above and below the TMD), $\varepsilon_r=4.58$ for bilayer graphene, and $\varepsilon_r=3.8$ for SiO$_2$.}
\begin{tabular}{ccccc|cc}
\hline
\hline
 & Substrate & Theory & Experiment & \multicolumn{3}{c}{Present paper} \\
 &  &  &  &  0 T & 10 T & 20 T\\
\hline
\hline
Mo$\text{S}_2$ & Vacuum & 551.4 [\onlinecite{theory}] & 570 [\onlinecite{MoS2exc2}] & 555.7 & 556.9 & 558.1 \\
 & & 526.5 [\onlinecite{theory2}] & & & & \\
 & & 555.0 [\onlinecite{analytic}] & & & & \\
 & hBN & 222.0 [\onlinecite{hBN}] & & 188.2 & 189.4 & 190.6 \\
\hline
MoS$\text{e}_2$ & Vacuum & 477.8 [\onlinecite{theory}] & & 486.7 & 487.8 & 488.8 \\
 & & 476.9 [\onlinecite{theory2}] & & & & \\
 & & 480.4 [\onlinecite{analytic}] & & & & \\
  & SiO$_2$ & & 590 [\onlinecite{MoWSe2exc}] & 291.4 & 292.5 & 293.5 \\
  & Bilayer & & 550 [\onlinecite{MoSe2exc}] & 261.7 & 262.8 & 263.8 \\
  & graphene & & 580 [\onlinecite{MoSe2exc2}] & & & \\
\hline
W$\text{S}_2$ & Vacuum & 519.1 [\onlinecite{theory}] & & 530.1 & 532.0 & 533.7 \\
 & & 509.8 [\onlinecite{theory2}] & & & & \\
 & & 523.5 [\onlinecite{analytic}] & & & \\
 & Si$\text{O}_2$ &  & 320 [\onlinecite{he}] & 289.2 & 291.0 & 292.7 \\
 & & & 312 [\onlinecite{perpfield}] & & & \\
 & & & 710 [\onlinecite{WS2exc}] & & & \\
 & & & 410 [\onlinecite{ws2dia}] & & & \\
\hline
WS$\text{e}_2$ & Vacuum & 466.7 [\onlinecite{theory}] & & 473.8 & 475.5 & 477.2 \\
 & & 456.4 [\onlinecite{theory2}] & & & \\
 & & 470.2 [\onlinecite{analytic}] & & & \\
 & Si$\text{O}_2$ & & 370 [\onlinecite{he}] & 265.2 & 267.1 & 268.7 \\
 & & & 720 [\onlinecite{MoWSe2exc}] & & & \\
 & & & 482 [\onlinecite{wse2dia}] & & & \\
 & & & 198 [\onlinecite{WSe2exctri}] & & & \\
\hline
\hline
\end{tabular}
\label{table:exctable}
\end{table}
\begin{table}
\centering
\caption{The same as Table \ref{table:exctable} but now for the negative trion binding energies (meV). We used $\varepsilon_r=3.8$ for SiO$_2$.}
\begin{tabular}{ccccc|cc}
\hline
\hline
 & Substrate & Theory & Experiment & \multicolumn{3}{c}{Present paper} \\
 & & & & 0 T & 10 T & 20 T \\
\hline
\hline
Mo$\text{S}_2$ & Vacuum & 33.8 [\onlinecite{theory}] & & 33.4 & 34.9 & 36.1 \\
 & & 32.0 [\onlinecite{theory2}] & & & & \\
 & & 33.7 [\onlinecite{analytic}] & & & & \\
 & & 32 [\onlinecite{falko}] & & & & \\
 & Si$\text{O}_2$ & & 18 [\onlinecite{mak3}] & 22.9 & 24.7 & 25.9 \\
\hline
MoS$\text{e}_2$ & Vacuum & 28.4 [\onlinecite{theory}] & & 27.7 & 29.3 & 30.4 \\
 & & 27.7 [\onlinecite{theory2}] & & & & \\
 & & 28.2 [\onlinecite{analytic}] & & & & \\
 & & 31 [\onlinecite{falko}] & & & & \\
 & Si$\text{O}_2$ & & 30 [\onlinecite{mak2}] & 20.3 & 21.7 & 22.8 \\
\hline
W$\text{S}_2$ & Vacuum & 34.0 [\onlinecite{theory}] & & 32.4 & 35.7 & 37.4 \\
 & & 33.1 [\onlinecite{theory2}] & & & & \\
 & & 33.8 [\onlinecite{analytic}] & & & & \\
 & & 31 [\onlinecite{falko}] & & & & \\
 & Si$\text{O}_2$ & & 30 [\onlinecite{WS2tri}] & 21.3 & 24.0 & 25.6 \\
 & & & 30 [\onlinecite{WS2tri2}] & & & \\
 & & & 18-45 [\onlinecite{WS2tri3}] & & & \\
 & & & 26 [\onlinecite{WS2tri4}] & & & \\
\hline
WS$\text{e}_2$ & Vacuum & 29.5 [\onlinecite{theory}] & & 28.8 & 31.3 & 32.7 \\
 & & 28.5 [\onlinecite{theory2}] & & & & \\
 & & 29.5 [\onlinecite{analytic}] & & & & \\
 & & 27 [\onlinecite{falko}] & & & & \\
 & Si$\text{O}_2$ & & 30 [\onlinecite{WeS2tri}] & 19.6 & 21.4 & 23.1 \\
 & & & 30 [\onlinecite{WSe2exctri}] & & & \\
\hline
\hline
\end{tabular}
\label{table:tritable}
\end{table}
\begin{table}
\centering
\caption{The same as Table \ref{table:exctable} but now for the biexciton binding energies (meV). We used $\varepsilon_r=3.12$ for Al$_2$O$_3$ and $\varepsilon_r=3.8$ for SiO$_2$.}
\begin{tabular}{ccccc|cc}
\hline
\hline
 & Substrate & Theory & Experiment & \multicolumn{3}{c}{Present paper} \\
 & & & & 0 T & 10 T & 20 T \\
\hline
\hline
Mo$\text{S}_2$ & Vacuum & 22.7 [\onlinecite{theory}] & & 19.0 & 22.3 & 23.5 \\
 & & 22.7 [\onlinecite{theory2}] & & & & \\
 & & 22.5 [\onlinecite{analytic}] & & & & \\
 & & 24 [\onlinecite{falko}] & & & & \\
 & $\text{Al}_2\text{O}_3$ & & 70 [\onlinecite{MoS2exc}] & 15.0 & 18.8 & 20.1 \\
\hline
MoS$\text{e}_2$ & Vacuum & 17.7 [\onlinecite{theory}] & & 15.5 & 17.6 & 19.3 \\
 & & 19.3 [\onlinecite{theory2}] & & & & \\
 & & 18.4 [\onlinecite{analytic}] & & & & \\
 & & 23 [\onlinecite{falko}] & & & & \\
 & $\text{Al}_2\text{O}_3$ & & 20 [\onlinecite{MoSe2biexc}] & 13.1 & 15.1 & 16.8 \\
\hline
W$\text{S}_2$ & Vacuum & 23.3 [\onlinecite{theory}] & & 19.5 & 22.4 & 25.0 \\
 & & 23.9 [\onlinecite{theory2}] & & & & \\
 & & 23.6 [\onlinecite{analytic}] & & & & \\
 & & 23 [\onlinecite{falko}] & & & & \\
 & Si$\text{O}_2$ & & 65 [\onlinecite{WS2tri}] & 14.5 & 17.1 & 19.7 \\
 & & & 69 [\onlinecite{WS2biexc}] & & & \\
\hline
WS$\text{e}_2$ & Vacuum & 20.0 [\onlinecite{theory}] & & 15.9 & 19.3 & 20.7 \\
 & & 20.7 [\onlinecite{theory2}] & & & & \\
 & & 20.2 [\onlinecite{analytic}] & & & & \\
 & & 20 [\onlinecite{falko}] & & & & \\
 & & 24.2 [\onlinecite{WSe2theory}] & & & & \\
 & Si$\text{O}_2$ & & 52 [\onlinecite{WeS2biexc}] & 12.5 & 15.3 & 17.6 \\
\hline
\hline
\end{tabular}
\label{table:biexctable}
\end{table}
\begin{table}
\centering
\caption{Exciton, trion, and biexciton diamagnetic shifts $\sigma$ ($\mu$eV T$^{-2}$) for different TMD materials on a SiO$_2$ substrate ($\varepsilon_r=3.8$) found by fitting the magnetic field dependence of the transition energy, which we compare with values calculated from the interparticle distances. Experimental results are shown for excitons. For WS$_2$ results are shown for both the $A$ and the $B$ excitons.}
\begin{tabular}{c|ccc|cc|cc}
\hline
\hline
 & \multicolumn{3}{c}{Exciton} & \multicolumn{2}{c}{Trion} & \multicolumn{2}{c}{Biexciton} \\
 \hline
 & Fit & Theory & Exper. & Fit & Theory & Fit & Theory \\
\hline
\hline
Mo$\text{S}_2$ & 0.033 & 0.035 & & 0.319 & 0.282 & 0.271 & 0.292 \\
\hline
MoS$\text{e}_2$ & 0.032 & 0.030 & & 0.263 & 0.286 & 0.302 & 0.287 \\
\hline
W$\text{S}_2$ $A$ & 0.080 & 0.080 & 0.9 [\onlinecite{perpfield}] & 0.756 & 0.700 & 0.633 & 0.744 \\
 & & & 0.32 [\onlinecite{ws2dia}] & & & & \\
W$\text{S}_2$ $B$ & 0.047 & 0.047 & 0.11 [\onlinecite{ws2dia}] & & & & \\
\hline
WS$\text{e}_2$ & 0.081 & 0.080 & 0.18 [\onlinecite{wse2dia}] & 0.992 & 0.709 & 0.849 & 0.790 \\
\hline
\hline
\end{tabular}
\label{table:diatable}
\end{table}

In Fig. \ref{fig:bigolfcont} the modulus squared of the biexciton wave function for a fixed electron and fixed holes, i.e. the conditional electron probability distribution, is shown for the previously used parameters $m_e=m_h=0.26m_0$ and $\kappa=1$. Notice that for $B=0$ T the other electron localizes predominantly around the fixed holes\cite{francoisT}. The presence of a strong magnetic field causes the localized regions around the fixed holes to merge, which is a manifestation of the fact that $\braket{r_{hh}}$ has decreased due to the magnetic field. \\

\subsection{Experimental systems}

In Tables \ref{table:exctable}, \ref{table:tritable}, and \ref{table:biexctable} we present the binding energies for excitons, negative trions, and biexcitons, respectively, for different materials, substrates, and magnetic field strengths and compare them with previous theoretical studies using ground-state diffusion Monte Carlo\cite{theory}, density functional theory and path-integral Monte Carlo\cite{theory2}, and the SVM\cite{analytic}, as well as experimental studies, for the case of zero magnetic field. No published magnetic field dependent results for the binding energy of excitonic systems in TMDs are available up to now. For these calculations we used the material constants given in Refs. [\onlinecite{berkelbach}] and [\onlinecite{theory2}], which we summarize in Table \ref{table:mattable}.

Our $B=0$ T results differ somewhat from the SVM results of Ref. [\onlinecite{analytic}] because we include the small difference between the effective electron and hole masses for the investigated TMDs whereas Kidd et al. [\onlinecite{analytic}] used equal electron and hole masses. When comparing the results from the different theoretical approaches with our results we see that the exciton and trion binding energies differ by at most 5\% while for the biexciton a 15-20\% smaller binding energy is obtained.

In experiments the TMD monolayers are usually placed on a substrate which enhances the dielectric screening in the TMD. This causes the binding energies of the excitonic systems to be lower than those of suspended TMDs in vacuum. From Table \ref{table:exctable} we notice that our calculation predicts an exciton binding energy which is about 5-30\% smaller than found experimentally, except for MoSe$_2$ as well as when comparing with Refs. [\onlinecite{WS2exc,MoWSe2exc}] where the disagreement is more than a factor of two. In Ref. [\onlinecite{wse2dia}] the high frequency value for the dielectric constant of SiO$_2$ is used instead of the static one, i.e. $\varepsilon_r=2.1$ instead of $\varepsilon_r=3.8$, which explains the larger binding energy. If we use $\varepsilon_r=2.1$ we find a binding energy of 363.6 meV for 0 T. As compared to experiment we underestimate the trion binding energy with about 30-35\%, except when comparing with Refs. [\onlinecite{mak3,WS2tri4}] where the disagreement is about 20\%.

In experiment one has found very large biexciton binding energies (see Table \ref{table:biexctable}) which can be up to a factor of 2-4 larger than predicted by theory. Notice that our theoretical biexciton binding energies are slightly smaller than those found from other theoretical approaches and therefore this large discrepancy between theory and experiment must be due to some other fundamental reason. It was argued by Kidd et al. [\onlinecite{analytic}] that this disagreement is due to a misinterpretation of the experimental results and that in experiment the particular biexciton peak is in fact the one from an excited state of the biexciton.

The binding energy of the different excitonic systems increases with magnetic field, where the increase is in general slightly larger at small fields, i.e. going from 0 T to 10 T, as compared to going from 10 T to 20 T. The materials with smaller effective electron and hole masses (W$\text{S}_2$ and W$\text{Se}_2$) exhibit a stronger magnetic field dependence than those with larger effective electron and hole masses (Mo$\text{S}_2$ and Mo$\text{Se}_2$). Furthermore, the magnetic field dependence is also more pronounced for trions and biexcitons as compared to that for excitons, which is a natural consequence of the fact that excitons are more strongly bound.

In Table \ref{table:diatable} we present the exciton, trion, and biexciton diamagnetic shifts, as determined from fitting the magnetic field dependence of the transition energy, and compare this with the theoretical value obtained from the calculated interparticle distance through Eq. \eqref{diashift}. Here we assumed equal electron and hole masses of 0.5$m_0$, 0.54$m_0$, 0.32$m_0$, and 0.34$m_0$ for MoS$_2$, MoSe$_2$, WS$_2$, and WSe$_2$, respectively. Only experimental results for excitons are shown in the table. To the best of our knowledge there are no experimental results for monolayers for biexcitons and only one for trions, i.e. $\sigma=5.7$ $\mu$eV T$^{-2}$ for WS$_2$ on a SiO$_2$ substrate\cite{perpfield}, which differs almost an order of magnitude from our results. The trion and biexciton diamagnetic shifts are comparable, whereas the corresponding exciton diamagnetic shift is almost an order of magnitude smaller. For excitons we find excellent agreement between the results obtained with the fit and the theory. For trions and biexcitons we find a relative discrepancy between the fit and the theory between 5\% and 28\%. This implies that the estimated size of the excitonic system will differ between 2.5\% and 15\% from the theoretical size. The agreement between the fit and the theory can be further improved by increasing the number of variational basis functions but this will lead to an exponential increase of the computation time. We also observe that the value of the diamagnetic shift for excitons, trion, and biexcitons depends strongly on the type of transition metal, whereas the type of chalcogen atom is of less importance.

\begin{figure}
\centering
\includegraphics[width=8.5cm]{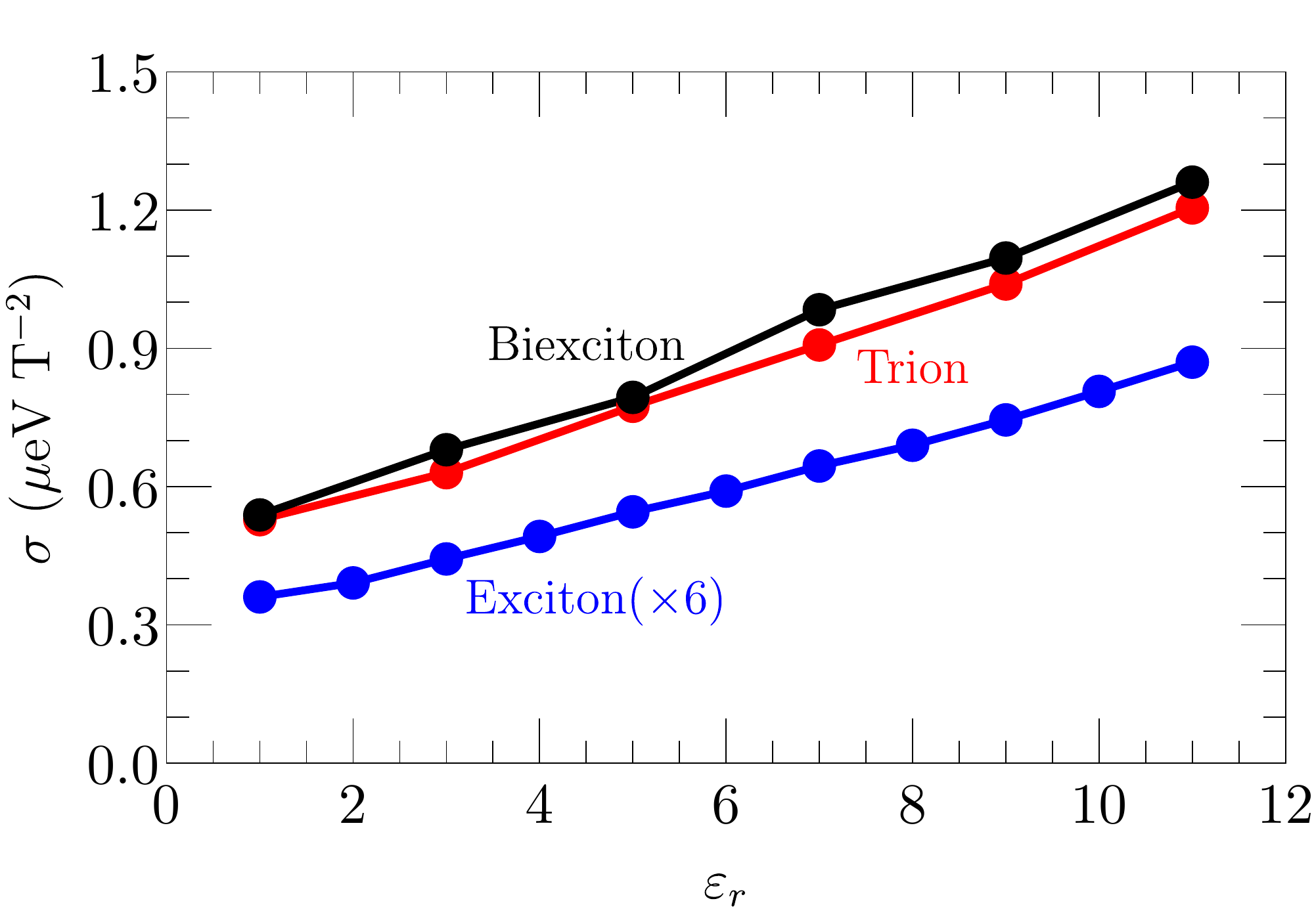}
\caption{(Color online) Exciton (blue), trion (red), and biexciton (black) diamagnetic shift obtained from Eq. \eqref{diashift} for WS$_2$ as a function of the dielectric constant of the substrate. The exciton results are scaled up by a factor 6.}
\label{fig:diaplot}
\end{figure}

For excitons, our results obtained from both the fit and Eq. \eqref{diashift} underestimate the experimental results by about a factor 4 for $A$ excitons in WS$_2$ and a factor 2 for $B$ excitons in WS$_2$ and for $A$ excitons in WSe$_2$. $A$ ($B$) excitons consist of a hole stemming from the upper (lower) valence band and have slightly different effective masses. For $B$ excitons in WS$_2$ we used $m_e=m_h=0.405m_0$. These experimental results depend significantly on the exact value of the dielectric constant of the substrate and in Ref. [\onlinecite{wse2dia}] the high frequency value for the dielectric constant of SiO$_2$ is used instead of the static one, i.e. $\varepsilon_r=2.1$ instead of $\varepsilon_r=3.8$. If we use $\varepsilon_r=2.1$ we find $\sigma=0.067$ $\mu$eV T$^{-2}$ and $\sigma=0.069$ $\mu$eV T$^{-2}$ through the fit and Eq. \eqref{diashift}, respectively. Furthermore, it is remarkable that the experimental results for these two materials differ by about a factor 2 while their effective charge carrier masses and screening lengths are very similar. We show the diamagnetic shift for WS$_2$ as a function of the dielectric constant of the substrate in Fig. \ref{fig:diaplot}, obtained from Eq. \eqref{diashift}. This shows that the diamagnetic shift increases approximately linearly with the dielectric constant of the substrate and that its value can be more than doubled as compared to the value in vacuum by choosing an appropriate substrate. The substrate dependence of the trion and biexciton diamagnetic shifts is stronger than that of the exciton diamagnetic shift. More specifically, we can fit the results for the three excitonic systems by a linear curve, i.e. $\sigma=a\varepsilon_r+b$ with $(a,b)=(0.0085,0.04887)$ $\mu$eV T$^{-2}$, $(a,b)=(0.0678,0.4408)$ $\mu$eV T$^{-2}$, and $(a,b)=(0.0721,0.4593)$ $\mu$eV T$^{-2}$ for excitons, trions, and biexcitons, respectively.

\section{Summary and conclusion}
\label{sec:Summary and conclusion}

In this paper, we used the stochastic variational method to investigate the binding energy and structural properties of excitons, trions, and biexcitons in 2D TMDs for different screening lengths and perpendicular magnetic field strengths, using the most simple isotropic effective mass approximation. For the exciton, we constructed a simplified variatonal method yielding results which are in good agreement with those found with the SVM.

We found that the binding energy of excitonic systems increases approximately linearly with magnetic field. For large screening lengths and high magnetic field strengths it is possible for the binding energy to deviate from this linear behavior. The magnetic field strength at which this deviation starts depends on how the magnetic length $l_B$ compares to the size of the excitonic system. When the magnetic length is smaller than the size of the excitonic system, the confinement will push the particles closer together and as such enhance their kinetic energy, which leads to a decrease in binding energy that adds up with the increase in binding energy stemming from the magnetic confinement to yield qualitatively different behavior. As trions and biexcitons are larger than excitons, their corresponding binding energies will deviate from the linear behavior at lower magnetic field strengths as compared to the corresponding exciton binding energy.

Furthermore, the binding energy of excitons decreases with the screening length, which can be understood since the dielectric screening decreases the short-range interactions. As a consequence, the dielectric screening leads to larger excitonic systems, which are therefore more sensitive to a perpendicular magnetic field.

The arguments above were confirmed by numerical calculations of average interparticle distances, interparticle distance probability distributions and moduli squared of wave functions, which clearly show that dielectric screening leads to an increase in size of the excitonic systems, whereas a perpendicular magnetic field leads to a decrease in size.

We also investigated the effect of a circular potential well on the binding energy of trions and biexcitons. We found that such a confinement potential also leads to an increase in binding energy and that this effect can be stronger than that of a perpendicular magnetic field.

Finally, we compared our results with those of other theoretical and experimental works in the absence of magnetic field. We found good agreement with other theoretical results for all three excitonic systems. Comparing with experimental results, we found reasonably good agreement for the exciton and for the trion. Our results disagree with the experimental data for biexcitons. It has been argued that this may be due to the fact that it is possible that in experiments excited states of the biexciton are observed. However, the current experimental works concerning the presence of external magnetic fields in 2D TMDs\cite{perpfield,perpabs1,perpabs2,perpabs3,perpabs4,ws2dia,wse2dia} do not specify a value for the binding energy in the presence of these fields and therefore a direct quantitative comparison for the binding energies is not possible. Therefore, we also looked at the exciton diamagnetic shifts and compared these with available experimental results and found that our results underestimate the experimental results, although this also depends on the exact value of the dielectric constant of the substrate. We also proposed to extend the concept of using the diamagnetic shift to get an estimate of the size of the excitonic system to trions and biexcitons and found that this estimate differs at most a factor 1.15 from the theoretical size.

\section{Acknowledgments}

This work was supported by the Research Foundation of Flanders (FWO-Vl) through an aspirant research grant for MVDD.

\end{document}